# Chiral asymmetry detected in a 2D array of permalloy square nanomagnets using circularly polarized X ray Resonant Magnetic Scattering

J. Díaz[1,2], P. Gargiani[3], C. Quirós[1,2], C. Redondo[4], R. Morales[5,6], L. M. Álvarez-Prado[1,2], J. I. Martín[1,2], A. Scholl[7], S. Ferrer[3], M. Vélez[1,2], and S. M. Valvidares[3]

[1]Depto. Física, Universidad de Oviedo, 33007 Oviedo, Spain.
[2]CINN (CSIC – Univ. de Oviedo), 33940 El Entrego, Spain.
[3]ALBA Synchrotron Light Source, Cerdanyola del Valles, 08290 Barcelona, Spain.
[4]Dept. of Chemical-Physics, Univ. of the Basque Country UPV/EHU, 48940 Leioa, Spain.
[5]Dept.of Chemical-Physics & BCMaterials, Univ. of the Basque Country UPV/EHU, 48940 Leioa, Spain.
[6]IKERBASQUE, Basque Foundation for Science, 48011 Bilbao, Spain.
[7]Advanced Light Source, Berkeley National Laboratory, 94720 Berkeley, USA

**Abstract**
The sensitivity of Circularly polarized X ray Resonant Magnetic Scattering (CXRMS) to chiral asymmetry has been demonstrated. The study was performed on a 2D array of Permalloy (Py) square nanomagnets of 700 nm lateral size arranged in a chess pattern, in a square lattice of 1000 nm lattice parameter. Previous X ray Magnetic Circular Dichroism Photoemission Electron microscopy (XMCD-PEEM) images on this sample showed the formation of vortices at remanence and a preference in their chiral state. The magnetic hysteresis loops of the array along the diagonal axis of the squares indicate a non-negligible and anisotropic interaction between vortices. The intensity of the magnetic scattering using circularly polarized light along one of the diagonal axes of the square magnets becomes asymmetric in intensity in the direction transversal to the incident plane at fields where the vortex states are formed. The asymmetry sign is inverted when the direction of the applied magnetic field is inverted. The result is the expected in the presence of an unbalanced chiral distribution. The effect is observed by CXRMS due to the interference between the charge scattering and the magnetic scattering.

1. **Introduction**

The characterization, modification and control of the chirality and/or polarity in 2D arrays of magnetic vortices is being the subject of intense research in magnetism due to the wide range of fundamental questions that this system pose and to the wealth of interesting applications derived from them [1]. They are envisioned to be an alternative for units of information storage in magnetic media due to their four possible vortex states [2]. And, more recently, they have been demonstrated to be an efficient source of spin waves, which is a basic device in modern spintronics based on magnons [3].
One important aspect to characterize in these systems is the level of degeneracy of their vortex states. These are determined by their chirality $c$ (in plane circulation of the magnetization, either left or right handed) and their polarity $p$ (out of plane magnetization at the core of the vortex, either pointing up or down the vortex plane). In perfectly symmetric systems, both kind of states are independent giving rise to four degenerated states. The breaking of this degeneracy is usually searched for practical purposes to favor, for instance, the presence of either a single chirality or a polarity state under certain conditions by modifying the shape and/or geometry of the magnets [4- 8]. Magnetostatic interaction between nanomagnets can be used to change the vortex state as well. When the magnets are close enough, the nucleation of the vortex states passes through a state where the flux closure involves more than one single magnet [9]. For instance, chains of vortices with alternate chiralities have been observed in square arrays of circular dots. [10].
The most used techniques to characterize vortex states are microscopies sensitive to the magnetization, like Magnetic Force Microscopy (MFM) [11], Lorentz transmission electron microscopy (TEM) [12] and X-ray microscopies such as Magnetic X-ray transmission microscopy (MXTM) [13, 14], and XMCD-PEEM [15]. MFM is the most extended technique since its relative simplicity makes it available at any laboratory. However, care must be taken to avoid the field from the tip to perturbing the imaged vortex states. This imposes certain limitations that affects the sizes and magnetizations of the observed nanomagnets and the environments under which the system is observed. In general, all these techniques are very good to capture the details of single vortex. Things become more difficult when those details must be checked in systems where collective states, like those described by Natali et al [10], are expected, or when statistical averaging over a large sampling of vortices is required.

X-ray techniques based on the reciprocal space are a good complement to the mentioned microscopic techniques. XRMS is a well stablish technique available at public synchrotron facilities which is especially sensitive to the detection of asymmetries and non collinear configurations in magnetic systems [16, 17]. It is considered the ideal technique to resolve and characterize spin spiral structures, such as skyrmions, the promising candidate for the advanced spintronics applications [18, 19]. It can be operated under the presence of magnetic and/or electrical fields, and it is element sensitive. Differently from other microscopic techniques like MFM, this technique is non-invasive, and it is especially sensitive to magnetic superorder [20, 21] and collective asymmetries. For instance, CXRMS has been used for stripe domains characterization [17], and to distinguish between Neél and Bloch walls, determining their chirality in magnetic domains of thin films with perpendicular magnetic anisotropy, i.e., with in plane and out of plane magnetization [22]. In those experiments, the magnetization asymmetries were detected by the intensity term related to solely the magnetic scattering. The present experiment explores the application of CXRMS in a periodic array of 2D square Py magnets, 700 nm wide, with only in-plane magnetization (except in the vortex core), that present vortex states and where the magnetostatic interaction between squares is relatively important. The special sensitivity of CXRMS to the chirality of the magnetic vortices in 2D magnets arrays is experimentally probed for the first time, and it is demonstrated to be due to the interference between the magnetic scattering and the charge scattering.

2. **Experiment**

**2.1 Sample Fabrication and characterization**

The 2D square array was produced by laser interference lithography [23]. A photosensitive stack made by a bottom antireflective coating and a polymeric resin was first spin coated on a Si substrate. The photosensitive coating was then illuminated by the interference pattern of a 325 nm wavelength laser. The regions exposed to the highest light intensity remain on the substrate while underexposed areas became soluble to the developer. Then, a layer of Py (80% Ni and 20 % Fe) of 25 nm thickness was deposited by magnetron sputtering from a single Py target. A protective capping layer of 2 nm of Al was deposited on top of the Py. The remaining polymer was removed by a chemical lift off process.

Figure 1 shows the hysteresis loops of the sample obtained by the longitudinal Kerr effect with the magnetic field applied in the direction parallel to the diagonals of the square magnets. The shape of the loop is the expected in presence of vortex states. There is a remnant magnetization which is about 2/3 of the magnetization at saturation. The coercive field is of about 2 mT. The critical field for vortex nucleation is similar for the two diagonal orientations, of about 3 mT. However, the vortex annihilation field varies with orientation, from 16 mT to 21 mT indicating a weak asymmetry between both orientations. According to Natali et al. [9], this difference corresponds to a weaker intermagnet magnetostatic interaction in the orientation with the higher annihilation field. This is likely caused by a larger gap distance between magnets and a shorter width of the magnets in that direction, as it can be deduced from the SEM image of the array in figure 2. To delimit the shape and size of the Py magnets, this image was obtained with a 30 keV incident beam and backscattered electron detection. The orientation chosen for the experiment was the one in which the interaction between the nanomagnets was weaker in the direction parallel to the incident beam. We define that direction being parallel to the X axis.

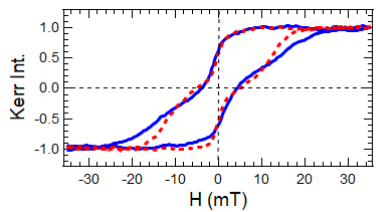

**Figure 1.** Magnetic hysteresis loop of the 2D square lattice array obtained by longitudinal Kerr effect with the magnetic field applied along the two diagonals of the square magnets (solid blue line and red broken line).

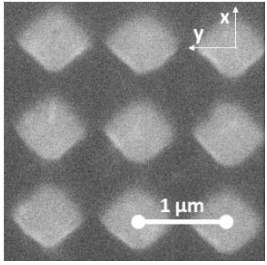

**Figure 2.** SEM images of the chess pattern array obtained using a 30 keV beam in back-scattered electron detection. The lattice parameter of the square lattice is 1 μm.

The presence of Landau closure domains in the nanomagnets in the absence of applied magnetic fields was confirmed by XMCD-PEEM. Figure 3 shows a PEEM image obtained at the L3 edge of Ni with the sample at remanence after completing a hysteresis loop along one of the diagonals axis of the squares. The image shows that there exists a preference in the chiral state of the vortices. The ratio observed in this image was of 20:8 between CCW against CW vortex. Images obtained over much larger areas (50 μm x 50 μm) confirm this result, with extended patches larger than the shown image in which the vortices were in a single chiral state.

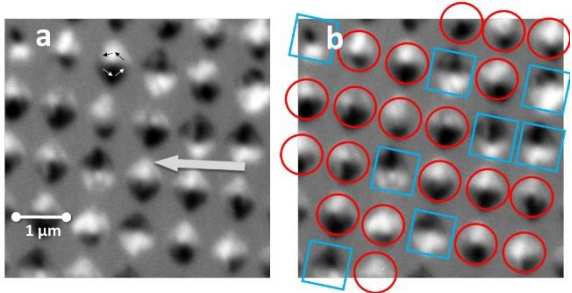

**Figure 3.** XMCD-PEEM image taken at the Ni L3 edge. The direction of the incident X ray beam is indicated with an arrow. Whiter regions indicate magnetic moments with the projected component parallel to the beam in the same direction of the incident beam. Darker regions indicate magnetic moments opposite to the direction of the beam. Image (b) is the same as image (a) with CCW chiral squares circled with red circles and CW chiral squares circled with blue squares.

## 2.2 X ray Experiment

The sample was inserted in a UHV chamber specially designed for soft X ray magnetic scattering experiments in BOREAS beamline at the ALBA synchrotron [24]. The X rays were 100% circularly polarized delivered from an undulator. The magnetic scattering measurements were performed at each applied magnetic field recording the scattered intensity at right and left circular polarization. The magnetic field was applied in the same orientation than the projected component of the incident beam on the plane of the sample, which we defined as the X axis. The geometry of the experiment is shown in Figure 4. We define the coordinate XYZ axes oriented with the plane XZ parallel to the plane of incidence of the X rays and the Y axis oriented perpendicular to this plane. The sample is in the XY plane. The magnetic field was generated by a specially designed superconductor magnet that guarantee a homogeneous magnetic field at the sample from 0 to 2 Tesla. The energy of the photons was tuned to the highest resonant magnetic scattering effect at the Ni L3 edge (851 eV). The plane of incidence of the X rays was parallel to the diagonal of the square nanomagnets. The chosen angle of incidence was 15°, covering a relative wide range of reciprocal space in the $q_x$ direction by the CCD camera. This angle of incidence is far from the critical angle and the charge scattering can be described using the simple Born approximation. The beam illuminated the sample after passing through a 50 μm diameter pinhole placed at 2 m from the sample. Data was recorded using a CCD detector located at 402 mm from the sample.

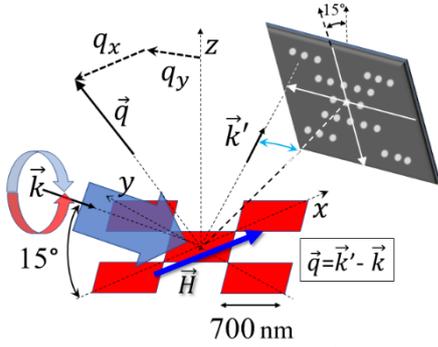

**Figure 4.** CXRMS experiment layout. The CCD screen, oriented perpendicular to the reflected beam, collects the scattered beam ($\vec{k}'$). The image is latter transformed to $\vec{q}$ coordinates by making $\vec{q} = \vec{k}' - \vec{k}$, where $\vec{k}$ is the wavevector of the incident beam.

## 3 Results
### 3.1 X ray Difraction

Figure 5 is an image of the diffraction pattern of the sample obtained by summing the intensity registered in the CCD detector for left and right circular polarizations, ($C^+ + C^-$), at the L$_3$ edge of Ni (851 eV photon energy) at a field of 2 T, when the sample is magnetically saturated and well above the vortex annihilation field. The CCD image has been converted to $\vec{q}$-space by making $\vec{q} = \vec{k}' - \vec{k}$, where $\vec{k}'$ and $\vec{k}$ are the scattered and incident X-ray wave vectors respectively. The specular reflection is at the point where $q_x$ and $q_y$ are 0 ($[q_x q_y] = [00]$, $q_z = 2k \sin \theta_i$). In the representation of the CCD detector of figure 4, the horizontal intensity variation can be directly mapped to $q_y$, whereas vertical variations come from the projection of $q_x$ and $q_z$ on the CCD plane. The variations in intensity related to $q_z$ are produced by the thickness of the magnets. For the range of $q_z$ scanned, these variations are relatively small. The images show only $q_x$ coordinates, which is where the variations in intensity due to the periodicity of the array and to its charge and magnetic structure are expected.

The $C^+ + C^-$ pattern remains constant independently of the applied magnetic field confirming its only-charge scattering contribution. The position of the peaks is that of a square lattice with a lattice parameter of 1000 nm. The modulation in intensity of the diffracted peaks differs in some aspects from the expected form factor of a perfect flat square, which is similar to an "X" with the highest intensity at the center ($[q_x q_y] = [00]$). The measured form factor shown in figure 5 (a) is more rounded at the corners, and the intensity is broadly distributed around the center. The pattern is symmetric along the perpendicular scattering direction ($q_y$). However, it is not symmetric in intensity in the $q_x$ axis, being collected more intensity in the region with $q_x < 0$. This is better shown in figures 5 (b) and 5 (c) which display intensity scans taken at the $[q_x 0]$ and $[0 q_y]$ directions of the diffraction pattern. The reflected peak ([00] peak) is barely more intense than the peaks at the [10] or [01] positions. The spread and asymmetric distribution of the scattered intensity could be due to deviations from the ideal flatness of the nanomagnets, and, possibly, to a distribution in the orientation of the surface of the different nanomagnets around a main orientation angle, like the mosaicity in crystalline materials. The estimated width in angle of these deviations is lower than 3° (50 mrad), the total angular width covered by the detector in the image of figure 5 (a).

The width of the diffracted peaks is affected by the 50 μm aperture and the 15° oblique angle incidence. It is smaller in the $[q_x 0]$ direction, where the transversal coherence length is larger, of the order of 5 unit cells. The coherence length in the $q_y$ direction is about four times smaller.

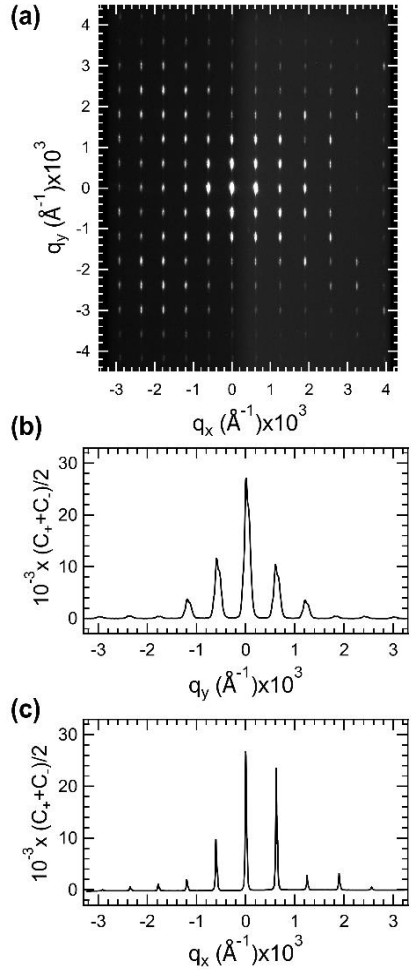

**Figure 5.** (a) Diffraction pattern of the 2D square magnets array obtained by summing the C+ and C- intensities measured at the CCD detector, (b) Intensity profile in the $[q_x 0]$ direction; (c) Intensity profile in the $[0\ q_y]$ direction. The specular reflection is at $[q_x q_y] = [00]$ ($q_z = 2k \sin \theta_i = 0{,}1$ Å$^{-1}$).

### 3.2 Magnetic Scattering

Figure 6 shows the patterns for the magnetic scattering, the $C^+ - C^-$ quantity, normalized to the charge scattering, i.e., to the $C^+ + C^-$ intensity, at different states of magnetization. In the positive saturation state, figure 6.1, the pattern looks very similar to the charge scattering pattern shown in Figure 5(a): with peaks arranged in a square lattice, symmetric in intensity in the $[0\ q_y]$ direction but asymmetric along the $q_x$ direction. This is the expected result since, in this case, charge and magnetic scatters are the same.

Along the full hysteresis loop, magnetic scattering peaks in figure 6 are always observed at the same positions as the charge scattering peaks in figure 5, independent of the magnetization state what discards the detection of magnetic superlattice order. At negative saturation, figure 6.3, the pattern is the same as in figure 6.1 but with inverted contrast since the magnetization direction is inverted with respect to the direction of the incident beam. A clear evolution of the XCRMS intensity is observed close to the remanence state where the magnetic scattering intensity is asymmetric in the $[0\ q_y]$ direction: as the magnetic field decreases from positive saturation down to negative saturation, XCRMS is more intense in the $q_y < 0$ region (figure 6.2). On the opposite, as the magnetic field increases from negative saturation up to positive saturation, XCRMS is more intense in the $q_y > 0$ region (figure 6.4).

Figure 7 shows in detail intensity profiles extracted from the CCD images in the $[0, q_y]$ direction at different magnetization states which shows more clearly the evolution in intensity of the diffracted peaks at different magnetization states. A set of positive peaks appears at +14 mT (figure 7.1) with the array still close to full positive saturation. Close to remanence, in the magnetization region where vortices states are expected, scattering peaks become weaker and are clearly asymmetric around zero (figures 7.2-7.3). As the field is reduced to -14 mT (figure 7.4), a regular pattern of negative peaks is observed corresponding to negative saturation. In the following, we will define as "positive/negative" asymmetry to $\Delta I = I_{-1} - I_{+1}$, the intensity difference between the [0,+1] and [0, -1] diffraction peaks. $\Delta I$ is negative in figures 7.2 and 7.3 but becomes negligible in the saturated states (figures 7.1 and 7.4). In the ascending field branch of the hysteresis loop (figures 7.5-7.6), the diffraction pattern becomes asymmetric again but with an opposite asymmetry sign. Finally, in figure 7.7, the diffraction pattern corresponding to the initial positive saturated state is recovered. That is, for a given branch of the hysteresis loop, the sign of the asymmetry stays constant and does not depend on the applied field direction: it is negative when the magnetization reverses from parallel to antiparallel to the incident X ray beam direction (descending field branch) and it is positive in the ascending field branch (magnetization direction changing from antiparallel to parallel to the incident beam direction).

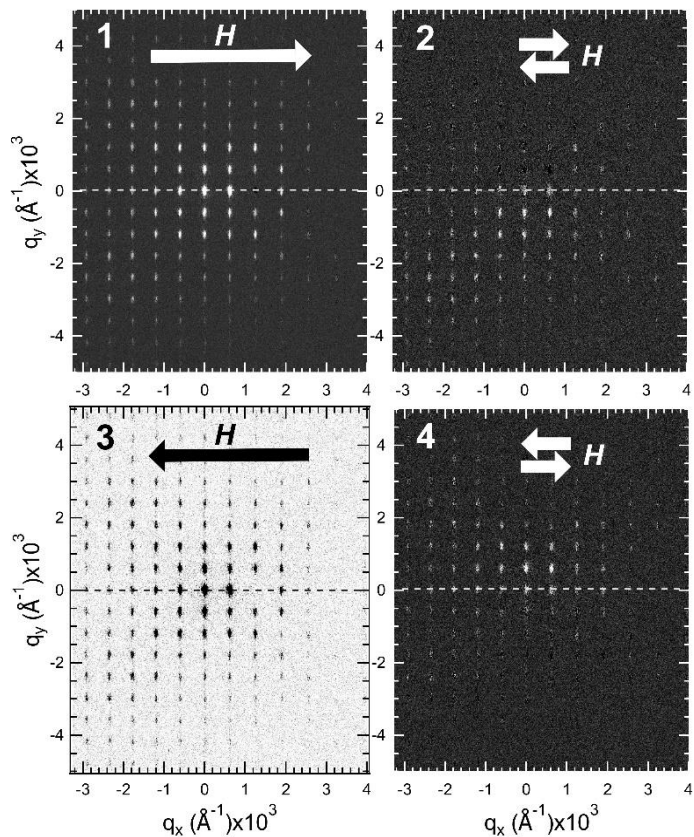

**Figure 6.** XRMS patterns ($C^+-C^-$) normalized to ($C^++C^-$) intensity and obtained at 4 different magnetization stages ordered in the time sequence: (1) positive saturation; (2) near cero magnetization from positive saturation; (3) negative saturation; (4) near zero magnetization from negative saturation. Note that the gray color of the background has been changed in image (3) to increase the contrast. The double arrow in graphs 2 and 4 indicates that the asymmetry in the pattern is independent of the magnetic field direction.

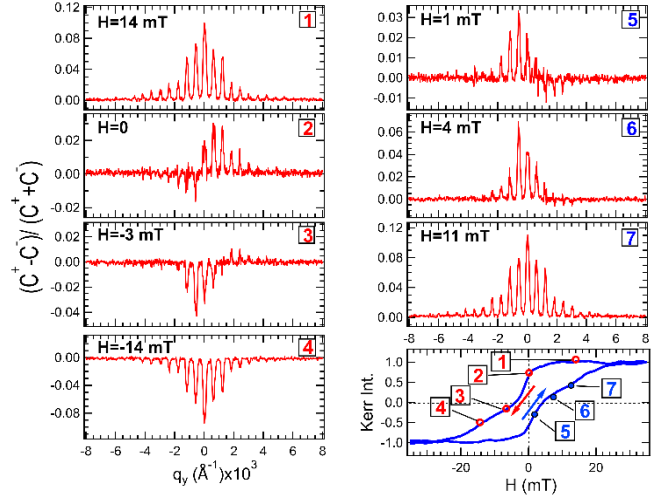

**Figure 7.** Sequence of CXRMS profiles extracted from the CCD images in the $[0, q_y]$ direction at different magnetic fields in a complete hysteresis cycle. The scans are displayed in a time sequence order.

## 4 Discussion

The intensity of the resonant magnetic scattering adds up to the originated by the charge. It is the sum of two contributions: one caused by the magnetic distribution at the scattered surface, and the other due to the interference between the charge and the magnetic scattering. According to [17], the expressions for these two intensities, written in $\vec{q}$-space for incident circularly polarized light, are the following:

$$I_m(\vec{q}) = |F^{(1)}|^2 Im[(\vec{k}' \cdot \vec{M}^*)(\vec{k}' \times \vec{k}) \cdot \vec{M}] P_3 \qquad (1)$$
$$I_i(\vec{q}) = Re\left[\bar{\rho}^* F^{(0)*} F^{(1)} \left(\vec{k} \cdot \vec{M} + (\vec{k}' \cdot \vec{k})(\vec{k}' \cdot \vec{M})\right)\right] P_3 \qquad (2)$$

$I_m(\vec{q})$ is the magnetic scattering and $I_i(\vec{q})$ is the charge-magnetic interference scattering. $\vec{k}'$ and $\vec{k}$ are the scattered and incident wave unit vectors respectively, where $\vec{q} = \vec{k}' - \vec{k}$. $\vec{M}(\vec{q})$ is the Fourier transform of the magnetization unit vector, i.e., the magnetization in $\vec{q}$-space. $F^{(0)}$ and $F^{(1)}$ are the Fourier transform of the charge and magnetic scattering factors, respectively. $\bar{\rho}(\vec{q})$ is the normalized Fourier transform of the charge distribution. $P_3$ is the Stokes parameter for circular polarization, which takes the values 1 ($C^+$) or -1 ($C^-$). Therefore, the sum of the intensities obtained with opposite polarizations cancel out the magnetic scattering, leaving only the intensity related to the scattering from the charge. Then, when the CXRMS resonant magnetic scattering, characterized by the $C^+ - C^-$ quantity, is normalized to the $C^+ + C^-$ intensity, as in the present work, the magnitude of the diffracted peaks yields the ratio of magnetic scattering against charge scattering at each point in the reciprocal space.

To understand the conditions for asymmetries in $q_y$ of expressions (1) and (2) as the experimentally observed, $\vec{M}(\vec{q})$ is decomposed in two components, one parallel ($\vec{M}_{\parallel}$) and the other perpendicular ($\vec{M}_{\perp}$) to the scattering plane. $\vec{M}_{\parallel}$ and $\vec{M}_{\perp}$ are related to the spatial magnetization configuration of the nanomagnets, given by the unit vector $\vec{m} = (m_x, m_y, m_z)$. From figure 4, $m_x$ and $m_y$ correspond to the in-plane parallel and perpendicular components of the magnetization with respect to the applied field, whereas $m_z$ comes from the magnetization of the vortex core. Thus, $m_z$ is only a minute fraction of the total magnetization of the permalloy squares and can be dismissed. The relationship between $(m_x, m_y)$ and $(\vec{M}_{\parallel}, \vec{M}_{\perp})$ depends on the orientation of the scattering plane relative to the sample. When $q_y = 0$, the scattering plane is the XZ plane so that $\vec{M}_{\parallel}$ can be associated to $m_x$ and $\vec{M}_{\perp}$ to $m_y$. Whenever $q_y$ is different from cero, the orientation of the scattering plane changes. Then, $\vec{M}_{\parallel}$ and $\vec{M}_{\perp}$ acquire additional $m_y$ and $m_x$ components, respectively. The magnitude of these components is proportional to the angle between the scattering plane and the one when $q_y = 0$. In the present case, this angle is very small, of the order of $\frac{q_y}{k_0}$ (about 20 mrad) and these additional components can be dismissed.

In a geometrically centrosymmetric structure like the squares, the real parts of $\vec{M}_\parallel$ and $\vec{M}_\perp$ ($Re\{\vec{M}_\parallel\}$ and $Re\{\vec{M}_\perp\}$) are even functions in $\vec{q}$ whereas the imaginary parts ($Im\{\vec{M}_\parallel\}$ and $Im\{\vec{M}_\perp\}$) are odd functions in $\vec{q}$. This means that the real part of each magnetization component is associated to the net magnetization in the corresponding direction whereas its imaginary part corresponds to the presence of an inversion symmetry axis, i.e., to the change of sign of the magnetization from one side to the other of the corresponding symmetry axis of the square magnet. Therefore, $Im(M_\parallel)$ and $Im(M_\perp)$ are odd in $q_y$ and $q_x$, respectively. The effect of these symmetries in the XCMRS signal is better understood if we rewrite expressions (1) and (2) in an alternative way:

$$I_m(\vec{q}) = \epsilon |F^{(1)}|^2 [Re(M_\parallel)Im(M_\perp) - Im(M_\parallel)Re(M_\perp)]P_3 \qquad (3)$$

The intensity $I_i(\vec{q})$ can be also decomposed in two terms, since the Fourier transform of the scattering coefficients $F^{(0)*}F^{(1)}$ are complex numbers. The resulting expression is an odd function of only the real and imaginary parts of $M_\parallel$:

$$I_i(\vec{q}) = \epsilon'[Re(\bar{\rho}^* F^{(0)*}F^{(1)}) \cdot Re\{M_\parallel\} - Im(\bar{\rho}^* F^{(0)*}F^{(1)}) \cdot Im\{M_\parallel\}]P_3 \qquad (4)$$

The terms $\epsilon$ and $\epsilon'$ depend on the incident and scattering angles and their value is close to 1. The coefficients $|F^{(1)}|^2$, $Re(F^{(0)*}F^{(1)})$ and $Im(F^{(0)*}F^{(1)})$ are all of the same order of magnitude. Their exact values depend on the exact photon energy used in the experiment.

When the sample is magnetically saturated along the $x$ axis, i.e. in a state of uniform magnetization with $m_x = 1$ and $m_y = 0$, $Im(M_\parallel)$ and $Im(M_\perp)$ are cero due to the spatial symmetry, $Re\{\vec{M}_\perp\}$ is cero since $m_y = 0$ and only $Re\{M_\parallel\}$ presents a finite value. Then, equations (3) and (4) imply that $I_m(\vec{q})$ is zero and the XCMRS signal comes only from $I_i(\vec{q})$.

At remanence, the square magnet should be in the vortex state with zero net magnetic moment, $m_x=0$ and $m_y=0$, so that $Re\{M_\parallel\} = Re\{\vec{M}_\perp\} = 0$. Expressions (3) and (4) show easily that $I_m(\vec{q}) = 0$ and $I_i(\vec{q}) \approx [Im(\bar{\rho}^* F^{(0)*}F^{(1)}) \cdot Im(M_\parallel)]P_3$. This demonstrates that the term $I_i(\vec{q})$ is odd in $q_y$ since it is proportional to $Im(M_\parallel)$, with opposite asymmetry sign for CW and CCW chiralities. Thus, the net asymmetry of $I_i(\vec{q})$ in $q_y$ seen in the experimental results of Figs. 5 and 6 can be the result of an unbalance in the proportion of CW/CCW chirality within the 2D array of magnetic vortices.

The analysis of $I_m(\vec{q})$ with eq. (3) shows other possible magnetic configuration that could produce a XCMRS signal asymmetric along $q_y$ but not along $q_x$, as experimentally observed. $I_m(\vec{q})$ will be odd in $q_y$ if $Re(M_\perp)$ is different from cero. Additionally, the chirality must be balanced, which is a necessary condition to avoid an asymmetry in $q_x$ from the $Im(M_\perp)$ term, since $Re(M_\parallel)$ is different from cero at any point of the hysteresis loop of the square magnet (except at remanence).

The comparison between the expected intensities of the terms $I_i(\vec{q})$ and $I_m(\vec{q})$ for the condition of antisymmetry in $q_y$ discards the later. Although their scattering coefficients might be of similar magnitude, the intensity of $I_m(\vec{q})$ should be lower since it depends on the magnitude of $Re(M_\perp)$, which is associated to $m_y$. This magnetization component should be small in an easy axis hysteresis loop as studied here. In a single vortex, the net $m_y$ is proportional to vortex core displacements parallel to the X axis. An estimation of magnetic scattering for a vortex core displacement of, for instance, a quarter of the lattice parameter yields $I_m(\vec{q})$ 20 times smaller than $I_i(\vec{q})$. This is much smaller than the ratio between the magnetic scattering intensity observed at the lowest magnetization state and in saturation (of about 1/3 in figure 7). Moreover, XMCD-PEEM images do not show clear displacements in the vortex core of the square magnets to evidence a $m_y$ component in their magnetization. Consequently, we can safely say that the observed asymmetry in the intensity along $q_y$ is due to an unbalanced chiral distribution, which is consistent with the PEEM images in figure 3.

Then, the changes in the magnetic scattering observed at different points of the hysteresis loop of the sample, displayed in figures 6 and 7, are mainly caused by the interference term $I_i(\vec{q})$. At saturation, the change in the sign of the contrast of figures 6.1 and 6.3 (or scan profiles 7.1 and 7.4), in which the magnetization is saturated at opposite directions, is explained because $\vec{M}_\parallel$ is Real and $I_i(\vec{q})$ is odd in $Re(M_\parallel)$. As the field decreases in magnitude and takes values below the vortex nucleation field, $Re(M_\parallel)$ becomes smaller and $Im(\vec{M}_\parallel)$, which is odd in $q_y$, becomes more important.

The sign of the asymmetry changes with the branch of the hysteresis loop, i.e., it changes with the direction of the applied field at which the vortices nucleate. This means that, since the observation direction (X ray wave vector incident direction) keeps fixed, the system keeps the same chirality all through the complete hysteresis loop. This happens even if the magnetization of the square magnets array is completely reset at the beginning of each hysteresis loop branch with fields much higher (2 T) than the annihilation fields (20 mT). This demonstrates that the chiral state of the array of square Py magnets is fixed for the chosen magnetic field orientation, i.e., that the sample presents chiral asymmetry.

**Conclusions**

The sensitivity of CXRMS to chiral asymmetry in 2D vortex arrays has been demonstrated and the conditions for their observation have been explained in detail. The effect is entirely due to the term of intensity related to the interference between the charge and the magnetic scattering. This contrasts with the observations made in other magnetic systems in which their out-of-plane magnetization component has a significant magnitude, whose asymmetries in their magnetization distribution were detected solely by the pure magnetic scattering term [17,22]. The demonstrated sensitivity of CXRMS to the chirality of 2D magnetic vortex opens the door to the study of more complicated systems where the potential capabilities of the technique, like non-invasive detection, element sensitivity, time resolved dynamics and/or magnetic superlattice order sensitivity can be exploited to the study of 2D magnet arrays even at smaller scales than the presented in this experiment.

**Acknowledgments**
This work has been supported by Spanish MINECO under grants FIS2013-45469
 and FIS2016-76058 (AEI/FEDER,EU)).